\documentclass[aps,pre,superscriptaddress]{revtex4}
\usepackage{graphics,graphicx,amsmath,amssymb,textcomp}
\usepackage{dcolumn}
\usepackage{bm}
\newcommand{\dan}{\affiliation{McMaster University, Department of Engineering Physics, Hamilton, Ontario, Canada L8S 4L7}}
\newcommand{\mike}{\affiliation{Scotiabank, Toronto, ON, Canada M5H 1H1}}
\newcommand{\rachid}{\affiliation{Physics\&Astronomy, University of Calgary, Calgary,
Alberta Canada, T2N 1N4}}
\newcommand{\ouyed}{\affiliation{
 Origins Institute,  McMaster
University, Hamilton, Ontario L8S 4M1, Canada}}

\begin{document}
\title{Pricing European Options with a Log Student's
{\it t}\thinspace-Distribution:\\ a Gosset Formula}

\author{Daniel T. Cassidy}%
\email{cassidy@mcmaster.ca}
\dan
\author{Michael J. Hamp}%
\email{mike_hamp@scotiacapital.com}
\mike
\author{Rachid Ouyed}%
\email{ouyed@phas.ucalgary.ca}
\rachid
\ouyed

\begin{abstract}
The distribution of the returns for a stock are not well described by a normal probability density function (pdf).  Student's \textit{t}-distributions, which have fat tails, are known to fit the distributions of the returns.  We present pricing of European call or put options using a log Student's \textit{t}-distribution, which we call a Gosset approach in honour of W.S. Gosset, the author behind the {\it nom de plume} Student.  The approach that we present can be used to price European options using other distributions and yields the Black-Scholes formula for returns described by a normal pdf. \newline \newline
Keywords: Econophysics; Financial risk; European options; Fat-tailed distributions; Student's $t$-distribution
\end{abstract} 

\maketitle
\section{Introduction}

	In the fall of 2008, many investors and analysts witnessed multiple \textquotedblleft once-in-a-lifetime\textquotedblright\space events in a single week, with disastrous consequences in many cases for the financial health of their portfolios.  The once-in-a-lifetime designation of the events arises from a calculation of the probabilities of events based on normal statistics.  The returns were more than several sigma beyond the expected return, and by normal statistics, thought to be impossible. 

	The difficulty lies in the use of a normal pdf to describe returns.  It is known that returns have \textquotedblleft fat tails\textquotedblright\space \cite{mandelbrot63,fama65}, but for mathematical convenience and perhaps force of habit, a normal pdf is often applied.  The celebrated Black-Scholes formula for pricing European options is based on several assumptions, one of which is that the returns are described by Brownian motion \cite{lax2006,bs73,Hull2006}.  The underlying pdf for Brownian motion is a normal pdf.

	In this paper we price European options using a log Student's $t$-distribution.  We call these formulae of the prices of European options Gosset formulae, in honour of the author behind the \textit{nom de plume} Student.  We present evidence that stock returns are fit by a Student's \textit{t}-distribution, in agreement with known results \cite{Blattberg74,jong00,Platen07,zhu2009}.  We then demonstrate the \textquotedblleft fat tails\textquotedblright\space of the \textit{t}-distribution.  One of the difficulties with the fat tails is that one of the integrals, which is required to price an option, diverges.  We present two similar approaches to handle successfully the divergence, and hence find a price for the option.  These approaches can be used to price options using pdf's other than the $t$-distribution.  However, we restrict our attention to the $t$-distribution and compare prices for options using the Gosset and Black-Scholes formulae.  In general, the Gosset formula yields prices for options that are higher than the prices found from the Black-Scholes formula.  This is not unexpected, as the Gosset formula gives greater weight to events in the tails.   

	The starting point for the pricing of the European options is the arbitrage theorem \cite{finetti1963,Health-Sudderth1972,Ross2007}.

\section{Arbitrage Theorem}

	Consider an experiment that returns $r_i(e_j)$ for a bet of unity on outcome $e_i$ of the $N$ possible outcomes $e_j$ of the experiment. In short, the arbitrage theorem states that either the mean return as calculated over the probability of the events $e_j$, $E\{r_i (e_j)\}$, is zero or there is a betting scheme that leads to a sure win.  The betting scheme may allow for negative, zero, and positive bets.  Puts and calls allow for negative, zero, and positive bets. 

	Let $e_i$ be $N$ disjoint events of which one and only one must occur.  Let $p_i$  be the probability of occurrence of event $e_i$ and \textbf{\textit{c }}$ = (c_1, c_2, ... c_N)$ be a vector of bets on the $N$ outcomes.  The bets $c_i$ may be zero, positive, or negative.  The gain for a wager of $c_i$ on the $i^{th}$ outcome given that event $e_j$ occurred is $g_i(e_j) = c_i \times r_i(e_j)$.  Note that in a typical bet,  $r_i(e_j) = -1$ if  $i\ne j$; the player forfeits the wager $c_i$ if event $i$ does not occur.  The mean gain for a wager on the $i^{th}$  event is $\bar{g}_i = c_i \times \bar{r}_i$ and is the expectation over the $e_j$, which is obtained as $\sum_j p_j \times g_i(e_j) = c_i \times \sum_j p_j \times r_i(e_j)$. 

	Following the example of de Finetti \cite{finetti1963}, we recognize the $\bar{g}_i$ as $N$ equations in $N$ unknowns $c_{i }$.  To ensure that no arbitrage exists, the average gain $\bar{g}_i = 0$.  This is only possible for arbitrary $c_i$ if $\bar{r}_i = 0$.  If $\bar{r}_i \ne 0$, then the $c_i$  can be selected to yield, on average, $\bar{g}_i > 0$.  Thus to ensure no sure win, $\bar{r}_i = E\{r_i(e_j)\} = 0$.  That this condition must hold for all $i$ is a statement that there are no lucky numbers or values for assets.  In the context of pricing an option on an asset, the events $e_i$  could be the final price of the asset falling in a range. 
 
	Ross \cite{Ross2007} shows how to price an option using the arbitrage theorem, assuming that the theorem can be generalized to handle functions rather than $N$ discrete outcomes.  If the cost of an option is $C$, then by the principle of no-arbitrage and the arbitrage theorem $C = E\{\rho_i\}$, where the return for a European call option is $\rho_i = (S_i - K, 0)^+$ and the return for a European put option is $\rho_i = (K - S_i, 0)^+$.  The notation $(K -S, 0)^+$ means the maximum value of $(K - S)$ and 0, with $K$ being the strike price and $S$ the value of the asset. 
 
	Ross \cite{Ross2007} obtains the Black-Scholes formula by assuming log-normal statistics for the value of the asset and making some assumptions regarding the time value of the option.  It is not necessary to solve the Black-Scholes partial differential equation to obtain the Black-Scholes formula for the price of an European option \cite{Ross2007,gastineau1975}. 
 
	We accept as a starting point that the arbitrage theorem can be generalized from $N$ discrete outcomes to hold for functions.  This we consider a reasonable starting point.  The probability functions and values of assets can always be quantized to have $N$ discrete ranges.  We price European options assuming log Student's {\it t}-distributions, the arbitrage theorem, and results from martingale theory.  The results from the martingale theory allow determination of the risk neutral price of the option. 
 
	First we demonstrate that the log Student's {\it t}-distribution is a reasonable distribution to use to price an option.  

\section{Student's $t$-distribution}

The \textquotedblleft fat-tail\textquotedblright property was one of the earliest observations of asset returns, with published reports dating back to the early 1960's \cite{mandelbrot63,fama65}.  When the Black-Scholes formula was developed in the early 1970's the shortcomings of the underlying normal assumption were identified and attempts were made to incorporate kurtosis, see for example \cite{jarrow1982} and references therein.  Also in the early 1970's the Student's {\it t}-distribution was first used to describe asset returns \cite{Blattberg74}.  There has been considerable work since 1974 comparing the {\it t}-distribution and other distributions to empirical data to determine which distribution best fits observations, see for example \cite{Platen07}, and references therein.  The {\it t}-distribution has been found to provide realistic fits to observed asset returns (particularly equities and equity indices) and in many cases provides better fits than other distributions \cite{Platen07}.

In Fig. 1 we show fits to the observed frequency of occurence of daily return data for the Dow Jones Industrial Average (DJIA) (left) and the S\&P 500 (right) equity indices.  Each plot shows the data with a best fit {\it t}-distribution and normal distribution.  The {\it t}-distribution fits the observed distribution well, particularly in the tails.  We have observed similar quality fits for the {\it t}-distribution to frequency of occurence of daily returns for individual equities.

\begin{figure}[h!]
\label{fig:fig1}
\includegraphics[scale=0.4]{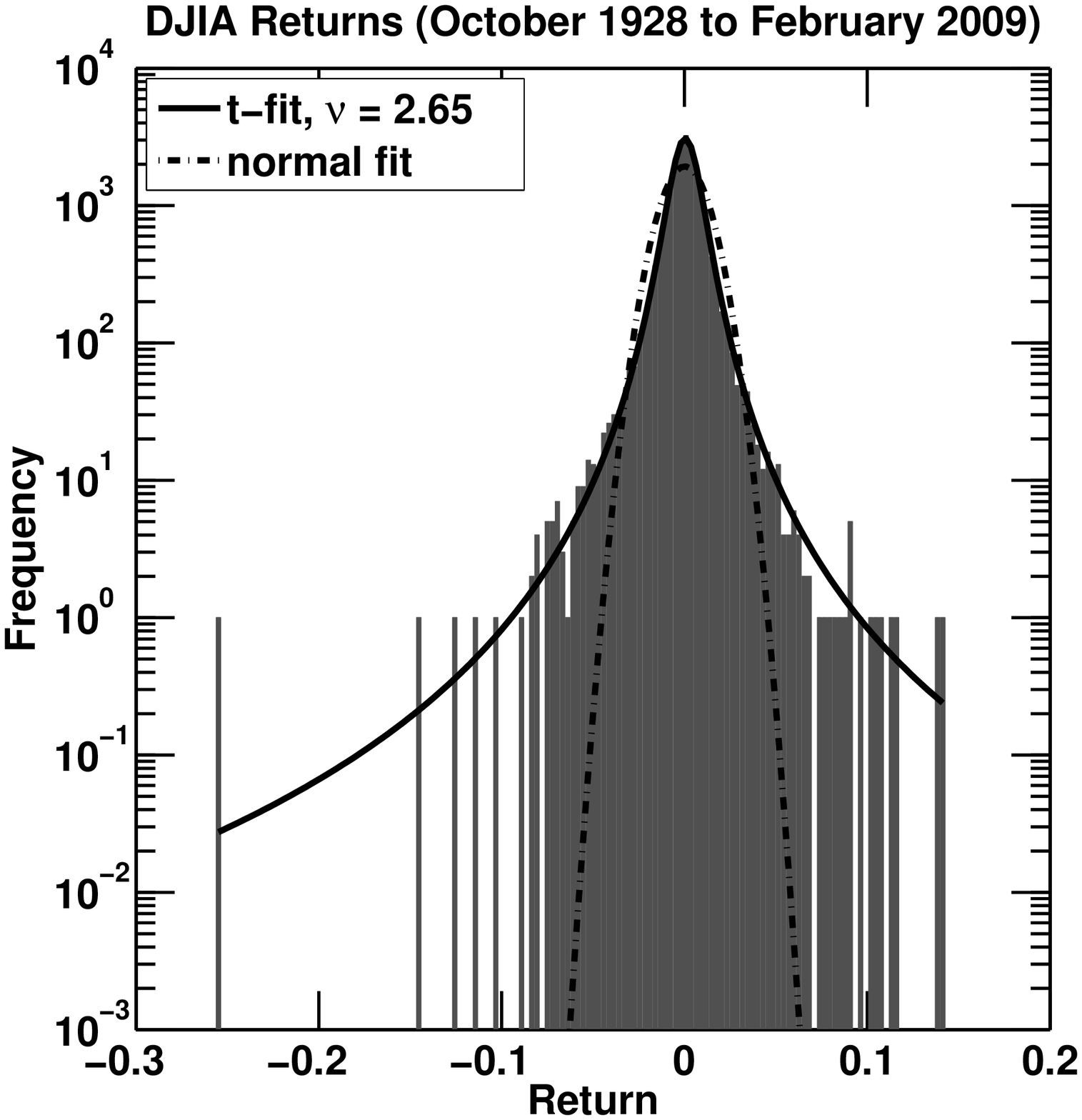}\includegraphics[scale=0.4]{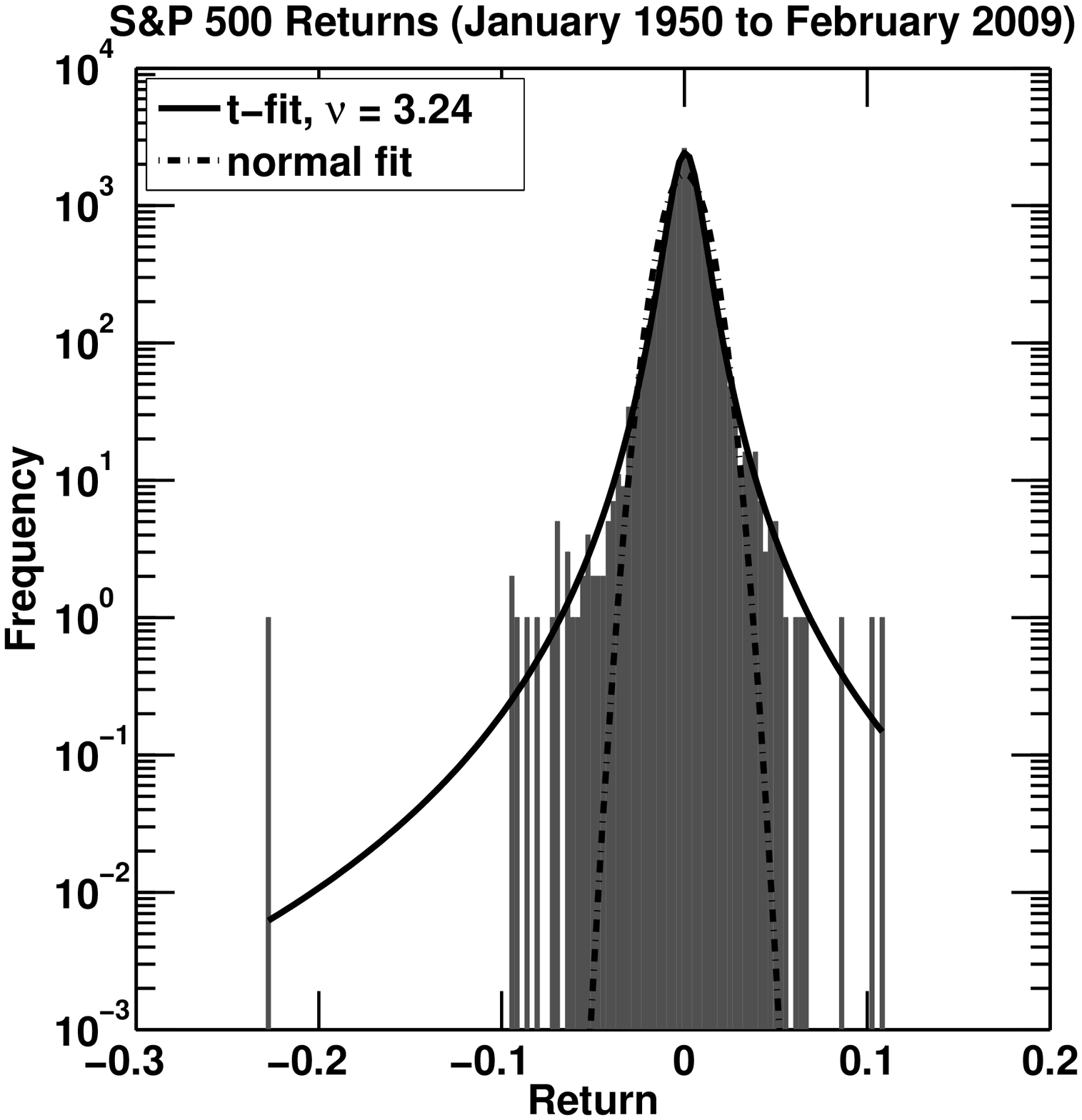}
\caption{Fit of Student's \textit{t}-distributions and normal distributions to returns for the DJIA and S\&P 500.
}
\end{figure}
	The fit paramaters with the associated uncertainties, some critical values, and sample descriptive statistics, such as the count of data points $N$, sample mean $\bar{r}$, and sample standard deviation $s$ are given in Table I.  The critical values $x_c(q)$ are the returns which solve $P\{\xi \leq x_c\} = q$.  Given that there are 21,186 data points for the DJIA, it is expected, using the best fit $t$-distribution, that only 2 returns will fall below -0.26.  This is in agreement with the data shown in Fig. 1. 
\setlength{\tabcolsep}{8pt}
\begin{table}[h!] 
\caption{Fit parameters, critical parameters, and simple descriptive statistics}
\begin{tabular}{c c c c c }\hline
   & \multicolumn{2}{ c }{DJIA} & \multicolumn{2}{ c }{S\&P 500}\\\hline 
parameter & normal & $t$ & normal & $t$\\\hline\\
location $\mu$ & $(1.679$\textpm$1.6)\times 10^{-4}$ & $(4.023$\textpm$1.1)\times 10^{-4}$  &$(2.660$\textpm$1.5)\times 10^{-4}$&$(4.270$\textpm$1.2)\times 10^{-4}$ \\
scale $\sigma$ & $(1.162$\textpm$0.011)\times 10^{-2}$ & $(1.157$\textpm$0.011)\times 10^{-2}$ & $(9.55$\textpm$0.11)\times 10^{-3}$ & $(8.71$\textpm$0.19)\times 10^{-3}$ \\
shape $\nu$ &   & 2.65 \textpm 0.11 &   &  3.24 \textpm {0.19} \\
 $|x_c(10^{-4})|$ & 0.043  & 0.26 & 0.036    &  0.19  \\
 $|x_c(10^{-3})|$ & 0.036  & 0.12  & 0.030  &  0.089 \\
 $|x_c(10^{-2})|$ & 0.027   & 0.053  & 0.022  &  0.040 \\
count & \multicolumn{2}{c} {$20,186$} &  \multicolumn{2}{c}{$14,870$} \\
$\bar{r}$ & \multicolumn{2}{c} {$1.679\times 10^{-4}$} &  \multicolumn{2}{c}{$2.660\times 10^{-4}$} \\
$s$ & \multicolumn{2}{c}{$0.01157$} &  \multicolumn{2}{c}{$0.00951$} \\ \hline
\end{tabular}
\end{table}
\vskip 0.05in
	Figure 2 is a plot of the value of the Student's {\it t}-distribution and the normal distribution as a function of the number of standard deviations from the true mean.  The fat tails of the Student's \textit{t}-distribution are evident.  Events that are 10 standard deviations from the true value are expected approximately $10^{-2}$\% of the time for the Student's {\it t}-distribution with a small number of degrees of freedom, but only $10^{-18}$\% of the time for a normal pdf.  For comparison, the values for $N(0, 1.5\sigma)$, a normal pdf with a mean of zero and a standard deviation $1.5\times$ greater than for the other curves is presented.  A volatility of $1.5\times$ the measured volatility is sometimes used to price options.  The enhanced volatility effectively gives the normal pdf \textquotedblleft fatter tails\textquotedblright.  It can be observed in Fig. 2 that the $N(0, 1.5\sigma)$ curve tracks the Student's {\it t}-distribution over a greater distance from the true mean. 

	One possible justification for the use of Student's $t$-distribution to describe the daily returns for an asset is as follows.  If the returns are assumed to be normally distributed with a given short term variance, the variance is taken to be a random variable that varies slowly in time, and $f_t(x)$ is the pdf for the return  at time some time $t$ in the future, then 
$f_t(x) = \int f_t(x|\sigma_t) \times pdf_{\sigma_t}(\sigma_t) d\sigma_t$ where $f_t(x|\sigma)$ is the pdf for the return given $\sigma$ and $pdf_{\sigma_t}(\sigma_t)$ is the pdf for $\sigma$ at time $t$.  Fits to the 22-day volatilities for the DJIA and for the S\&P 500 show that a gamma distribution fits the 22 day volatilities.  A gamma distribution with shape parameter $\nu$ is chi-squared with $2\nu$ degrees of freedom \cite{tref}, and a normal distribution with a variance that is chi-squared distributed is a Student's $t$-distribution.  A Student's $t$-distribution for the pdf of the daily returns is consistent with the data shown in Fig. 1.   

	The variance of a Student's $t$-distribution is $\nu/(\nu-2)$ \cite{tref}.  As $\nu$ tends to infinity, the $t$-distribution tends to a standard normal pdf, which has a variance of one.  The $\nu/(\nu-2)$ enhancement accounts for the range of values that the variance might take.  If it is known that the variance does not change in time (i.e, $pdf_{\sigma_t}(\sigma_t) = \delta(\sigma_t - \sigma_0)$ where $\delta(x)$ is a Dirac delta function), then the Gosset formula presented in this paper reduces to the Black-Scholes formula.  In this respect the Gosset formula extends the Black-Scholes formula by removing the condition that the variance is constant in time.  Under this interpretation, the shape parameter $\nu$ is a measure of the belief that the volatility will not change.  A large $\nu$ means that one believes that the volatility will not change in time.  A small $\nu$ means that one believes that there is a broad range of values that the volatility might take.  If the volatility is determined as the standard deviation of $N$ independent returns that are normally distributed with an unknown variance, then the maximum value of $\nu$ should equal $N-1$, which is the number of degrees of freedom in a determination of a sample standard deviation.  This maximum value of $\nu = N-1$ takes into account the fact that the sample standard deviation is a random variable and has an uncertainty associated with it.   

	Figure 3 is a plot of the relative areas in the tails for the same distributions presented in Fig. 2.  The curves are normalised by the area in the tail of a normal pdf.  Figure 3 gives the probability of an event $x$ for $x > t$, relative to the probability of the event $x$ for $x > t$ for a normal pdf.  	The probability of an event lying $> 10$ standard deviations from the mean is $10^{20}$ time mores likely for a process that follows a Student's {\it t}-distribution with 3 and 5 degrees of freedom than for a process that follows a normal pdf. 

\begin{figure}[h]
\begin{minipage}[t]{0.45\linewidth}
\centering
\includegraphics[scale=0.4]{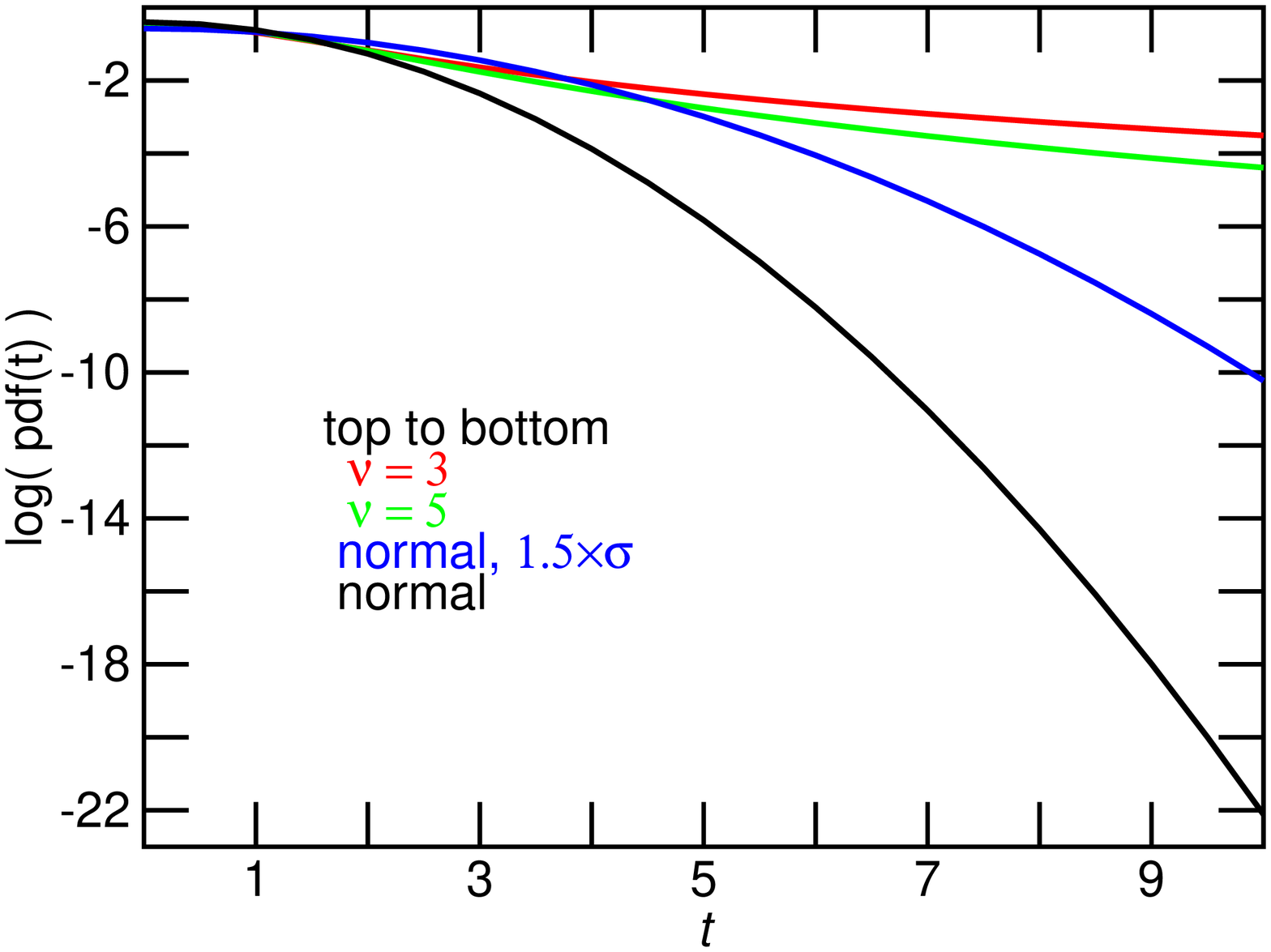}
\caption{Comparison of Student's \textit{t}-distributions and normal pdfs.}
\label{fig:fig2}
\end{minipage}
\hspace{0.6cm}
\begin{minipage}[t]{0.45\linewidth}
\centering
\includegraphics[scale=.4]{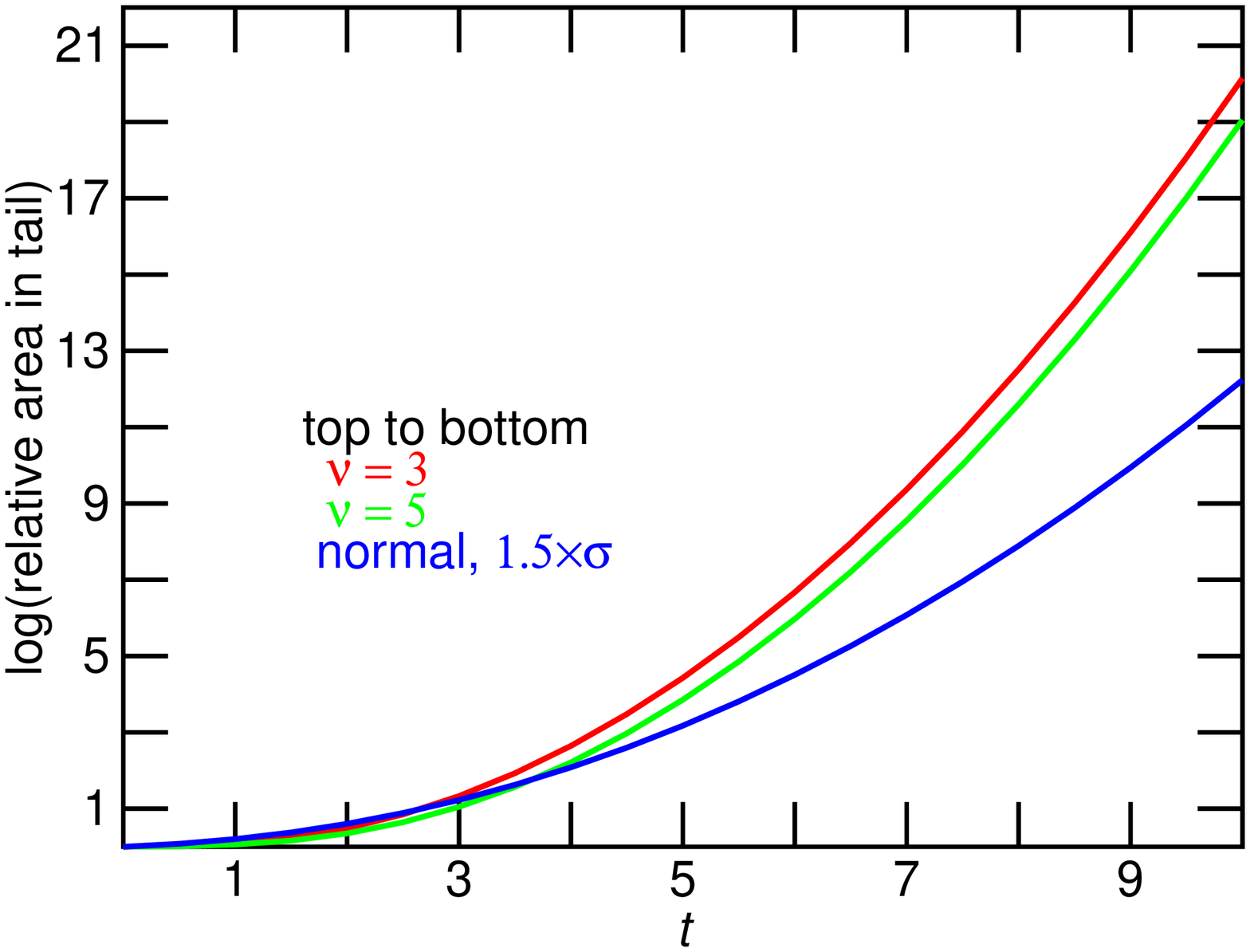}
\caption{Normalised areas in the tails for Student's \textit{t}-distributions with 3 and 5 degrees of freedom and for a normal distribution with $1.5\times$ the volatility, each normalised by the area in the tail of a normal pdf.}
\label{fig:fig3}
\end{minipage}
\end{figure}

\section{Pricing the Option}

	Le $S_t$ be the price of a stock at time $t, t > 0$.  Let $K_T$ be the strike price at time  $T$, where $T$ is the time when the option expires. 

	Let $S_t = A_t \exp( \sigma_{t }\xi )$ be the value of the stock where $\xi$ is a random variable. 

	The cost of a European call option, calculated at the time of expiration $T$, is $C_T = E\{(S_T - K_T)^+\} = \int pdf(S_T)\times(S_T - K_T)^+   dS_T$, which is the expectation of the maximum value of $\{S_T - K_{T }, 0\}$.  The desired quantity is $C_{0 }$, which is the value of the
option at time $t = 0$.  The desired quantity can be obtained from the expected time value of money.  If $r(t)$ is the risk free rate, then $C_0 = E\{C_T \times \exp(\int r(t)\times t dt)\} = C_T \times \exp(-r\times T)$ when the risk free rate is assumed to be time independent.  This is
a standard assumption in the derivation of the Black-Scholes formula \cite{bs73,Hull2006}. 

	If $\xi$ is normally distributed, then $S_t$ follows a log normal distribution and the price for the option follows the Black-Scholes formula  \cite{bs73,Hull2006}. 

	If $\xi$ follows a Student's \textit{t}-distribution, then the pdf for $\xi$ for $\nu$ degrees of freedom is
\begin{equation}
\label{eq:1}
pdf(\xi) d\xi = \frac{\Gamma(\frac{\nu+1}{2})}{\Gamma(\frac{\nu}{2})\sqrt{\pi \nu}}
\times \frac{d\xi}{\left(1+\frac{\xi^2}{\nu}\right)^{\frac{\nu+1}{2}}}\ = \Lambda(\nu) \times \frac{d\xi}{\left(1+\frac{\xi^2}{\nu}\right)^{\frac{\nu+1}{2}}}\ ,
\end{equation}
which can be written in terms of $S_t = A_t \exp(\sigma_t \xi)$ as
\begin{equation}
\label{eq:2}
pdf(S_t) dS_t =  \frac{\Lambda(\nu)}{\left(1+\frac{(\ln(S_t/A_t))^2}{\sigma_t^2\times \nu}\right)^{\frac{\nu+1}{2}}}
\times \frac{dS_t}{\sigma_t\times S_t}\ .
\end{equation}
	The following integrals are needed to evaluate the price of a call option at the time of expiration, $C_{T }$:

\begin{eqnarray}
\label{eq:3}
\int_{K_T}^{\infty} pdf(S_T)\times K_T\times dS_T  
&=& \int_{K_T}^{\infty}  \frac{\Lambda(\nu) \times K_T}{\left(1+\frac{(\ln(S_T/A_T))^2}{\sigma_T^2\times \nu}\right)^{\frac{\nu+1}{2}}}
\times \frac{dS_T}{\sigma_T\times S_T} \\\nonumber
&=& \int_{\frac{\ln(K_T/A_T)}{\sigma_T}}^{\infty}  \frac{\Lambda(\nu) \times K_T}{\left(1+\frac{\xi^2}{\nu}\right)^{\frac{\nu+1}{2}}}\times d\xi\ ;
\end{eqnarray}
\begin{eqnarray}
\label{eq:4}
\int_{K_T}^{\infty} pdf(S_T)\times S_T\times dS_T 
&=& \int_{S_T}^{\infty}   \frac{ \Lambda(\nu) \times K_T}{\left(1+\frac{(\ln(S_T/A_T))^2}{\sigma_T^2\times \nu}\right)^{\frac{\nu+1}{2}}}
\times \frac{dS_T}{\sigma_T\times S_T} \\\nonumber
&=& \int_{\frac{\ln(K_T/A_T)}{\sigma_T}}^{\infty} \Lambda(\nu) \frac{A_T\exp(\sigma_T\xi)}{\left(1+\frac{\xi^2}{\nu}\right)^{\frac{\nu+1}{2}}}\times d\xi\ .
\end{eqnarray}
Note that $A_T$ is a function of $S_0 , r, T, x_c(p)$, and $Z$, and that these variables are not functions of the random variable $\xi$ (see Eq. (6) and the accompanying text). 

	The integral in Eq.(4), which is for a call option, will be infinite for small $\nu$; the exponential term in the numerator will dominate the denominator.  Viable and physical results are obtained by restricting the upper limit of integration (i.e., truncate the pdf) or by restricting the value of the asset $S_T$ (i.e., cap the value).  It is not physical for the value of the asset $S_T$ to approach infinity. 

	For a put option this problem does not directly arise as $0 \le S_T \le K_T$ and thus the integrals are evaluated from $-\infty$  to $\ln(K_{T }/A_{T })/\sigma_{T }$.  However, the average price of the asset is needed to price an option, and thus the pricing of a put option suffers also from an integral that diverges. 

	The integral that diverges is Eq. (4).  If the value of the asset is restricted to be $\le A_{T }\exp(\sigma_{T }x_{c })$, then the offending integral, Eq. (4), can be rewritten as
\begin{eqnarray}
\label{eq:5}
\int_{K_T}^{\infty}  pdf(S_T)\times S_T\times dS_T  
&=&  \int_{\frac{\ln(K_T/A_T)}{\sigma_T}}^{\infty} \Lambda(\nu)
\times \frac{A_T\exp(\sigma_T\xi)}{\left(1+\frac{\xi^2}{\nu}\right)^{\frac{\nu+1}{2}}}\times d\xi \\\nonumber
&\ge& \int_{\frac{\ln(K_T/A_T)}{\sigma_T}}^{x_c} \Lambda(\nu)
\times \frac{A_T\exp(\sigma_T\xi)}{\left(1+\frac{\xi^2}{\nu}\right)^{\frac{\nu+1}{2}}}\times d\xi \\\nonumber  
&+& A_T \times \exp(\sigma_T x_c) \times  \int_{x_c}^{\infty} \Lambda(\nu)
\times \frac{d\xi}{\left(1+\frac{\xi^2}{\nu}\right)^{\frac{\nu+1}{2}}} .
\end{eqnarray}
	The integral in the last line of Eq.(5) is the probability $q = (1 - p)$ that the stock price $S_T = A_{T }\exp(\sigma_{T }\xi)$ is greater than some critical value $x_{c }(p)$, which we abbreviate to $x_{c }$.  The last line in Eq. (5) gives the expected value of the cost of capping the value of the asset to $A_{T }\exp(\sigma_{T }x_c)$. 

	In pricing a call option, one could also truncate the pdf by setting the last line of Eq. (5) to zero.  That is, one assumes that the pdf is zero for values greater than the critical value $x_{c }$.  Truncation sets an upper limit on the value of the stock:  $S_T\le x_{c }$.  It is necessary to divide the truncated pdf by $p$ to maintain the normalisation of the truncated pdf.  This follows from the definition of conditional
probability:  $P\{A \cap B\} = P\{A \cap B\}/P\{B\}$.  In this case, $B := 0 \le S_T \le S_{max} , A := S_{T }, A \cap B = S_T$ for $0 \le S_T \le S_{max}$ and equals the null set otherwise.  Thus $P\{S_T \vert 0 \le  S_T \le S_{max}\} = P\{S_T \}/P\{0 \le S_T \le S_{max}\} = P\{S_T\}/p$ if $0 \le S_T \le S_{max}$ and $= 0$ otherwise. 

	Table II gives critical values and the maximum increase in stock price $= \exp(\sigma_{T }x_c)$ for a normal pdf and a Student's {\it t}-distribution with $\nu = 5$ degrees of freedom, volatility $\sigma_T = 0.40$, and various value of $p$. 

	The effect of the fat tails of the Student's \textit{t}-distribution are evident.  For a confidence level of 99.99\%, the critical value for a Student's {\it t}-distribution with 5 degrees of freedom is 9.678 and the stock price has increased at most $47.992\times$, a 4 798\% increase. For a normal pdf, the critical value is 3.719 and there is a 99.99\% chance that the stock price has increased at most $4.426\times$, a 442\% increase.  The other interpretation is that there is only a 0.01\% chance that the stock price has increased by $> 47.992\times$ under the Student's {\it t}-distribution or $> 4.426\times$ under the normal distribution. 

	A confidence level of $p = 0.9999$ (i.e., 99.99\%) seems unrealistic.  This confidence level is one part in 10,000 and allows for increases of $\exp(\sigma_T x_x)$ in the price of the asset that may be physically impossible to achieve (c.f. Table III).  If the time period is one year, then once in 10,000 years is too large an extrapolation to be meaningful.  Even $p = 0.999$ seems unrealistic.  The best way to price options may be to pick an upper bound $x_c = \ln(S_{max})/\sigma_T$ (or confidence level $p$) for the price of the asset based on insight and agreement between parties, and then calculate the price using a capped distribution. 

	Although a confidence level $p = 0.9999$ might be unrealistic or unphysical in terms of expected increase in the stock price and extrapolation, the results presented here show that the methods allow a price for the option to be calculated.  The maximum increase in value of the asset is a function of the number of degrees of freedom and the assumed confidence level $p$ or critical value $x_{c }(p)$.  Table III presents the critical values for different values of $p$ and degrees of freedom $\nu$.  Note that a Student's {\it t}-distribution with $\nu =\infty$ yields a normal pdf and that a {\it t}-distribution with  $\nu \ge  40$ is often considered to be a normal pdf. 

	The variable $A_T$ is required to price the option.  This value is obtained from the assumption that an option is a fair wager (i.e., a martingale), as shown next. 
\setlength{\tabcolsep}{8pt}
\begin{table}[h!]
\caption{$S_t/A_t$ at $x_c(p)$ for volatility = 0.4 and selected $p.$}
\begin{tabular}{c| c c| c c }\hline
   & \multicolumn{2}{ c }{Student's {\it t}, $\nu=5$} & \multicolumn{2}{|c}{normal}\\\hline
$p$ & $x_c(p)$ & $\exp(\sigma_T x_c)$ & $x_c(p)$ & $\exp(\sigma_T x_c)$\\\hline
0.9   & 1.476      & 1.805                              &   1.282     & 1.670 \\
0.95 &2.015 & 2.239 &1.645 &1.931\\
0.99 &3.365 &3.842 &2.326 &2.536\\
0.995 &4.032 &5.018 &2.576 &2.801\\
0.999 &5.893 &10.56 &3.090 &3.442\\
0.9999 & 9.678 &47.99 &3.719 &4.428\\\hline
\end{tabular}
\end{table}
\vskip 0.05in
\begin{table}[h!]
\caption{Critical values $x_{c }(p)$ for various values of $p$ and $\nu$.}
\begin{tabular}{c|c c c c c}\hline
 $\nu$   &   $p=0.90$ &   $p=0.95$&   $p=0.99$&   $p=0.999$&   $p=0.9999$\\\hline
3 &1.638 &2.353 &4.541 &10.21 &22.20 \\
4 &1.533 &2.132 &3.747 &7.173 &13.03 \\
6 &1.440 &1.943 &3.143 &5.208 &8.025 \\
40 &1.303 &1.684 &2.423 &3.307 &4.094 \\
$\infty$ & 1.282 &1.645 &2.326 &3.090 &3.719 \\\hline
\end{tabular}
\end{table}
\vskip 0.05in
\section{Martingales, Doob Decomposition, and Risk Neutral Pricing}

	A martingale is a mathematical model for a fair bet\cite{Walsh2009}.  A fair bet can be defined as one for which the conditional expectation (i.e., given all the knowledge at the time of the bet) of the payout of the bet should be zero. 

	A martingale is defined such that $E\{X_{n +1}\vert F_n\} = X_n$ for all $n$ where $F_n$ is the knowledge (history) up to time $n$ and the knowledge increases as $n$ increases.  The  $X_n$ form a stochastic process.  A martingale is non-anticipating because the outcome of a martingale depends only on the past and not on the future. 

	A \textbf{sub}martingale is defined such that $X_n \le E\{X_{n+1}\vert F_n\}$ and a \textbf{super}martingale is defined such that $X_n \ge  E\{X_{n+1}\vert F_n\}$.  A submartingale represents a favourable process (the expected outcome is favourable) and a supermatingale represents an
unfavourable process (like a betting at a casino).  Casinos offer submartingales (the patrons play supermartingales) and financiers aspire to offer or to play submartingales. 

	If $X_n$ is a martingale, then $\exp(X_n), \exp(-X_n), \vert X_n\vert $, and $X_n^2$ are all submartingales. 

	The Doob decomposition shows that any submartingale $X_n$ is the sum of a martingale $M_n$ plus an increasing process, $X_n = M_n + A_n$, where $A_n \le A_{n+1}$ and  $A_n$ depends on the knowledge to $n-1, F_{n -1}$ \cite{Walsh2009}. 

	Let $S_t$ be the spot price at time $t$ for a stock.  $S_t = S_0 \exp(\alpha \times t)$ where $\alpha$ is the expected return on a non-dividend paying stock.  $\alpha$ is from a capital asset pricing model (CAPM) or equivalent \cite{Hull2006}.  $(\alpha - r)$ is called the risk premium and $\alpha = r + (\alpha-r)$ is the expected return on the stock, which equals the compensation for the time value of the money invested in the stock plus the compensation for the risk of holding the stock.  The time value is separated out since the value of the option $C_T$ is calculated at time $T$, but the value at time 0, $C_0$, is desired, and $C_0 = C_T \times \exp(-r\times T)$, assuming the risk free rate $r$ is known and constant in time. 

	The Doob decomposition is used to create a martingale for pricing options on the stock.  The drift of the stock is subtracted from the process.  This procedure of subtracting the drift from the process also leads to risk neutral pricing, where one is able to price an option without explicit knowledge of the risk premium. 

	The average value of a stock at time $t = 0$ when we wish to price an option is $S_0$.  The average price of the stock at some time later is $E\{S_t\} = E\{A_{t }\exp(\sigma_{t }\xi)\}$ where $\xi$ is a random variable and the drift is contained in $A_{t }$.  For a martingale (a fair wager), $E\{S_t\} = S_0\times exp(r\times t)$.  One may use the Doob decomposition to offset the drift and require $E\{A_{t }\exp(\sigma_{t }\xi)\} = \{A^{'}_{t }\exp(\sigma_{t }(\xi - \xi_0))\} = S_0\times \exp(r \times t)$. This is equivalent to requiring that the pdf for $S_t$ should be centred about the drift owing to the risk premium.  The $\exp(r\times t)$ term takes into account the time value of money and allows for comparison of values at two different points in time.

	It is perhaps simpler to find the normalization $A^{'}_{t }\exp(-\sigma_{t }\xi_0)$ that satisfies the martingale condition than to shift the pdf.  For a capped distribution,
\begin{eqnarray}
\label{eq:6}
\int_{0}^{\infty} pdf(S_T)\times S_T\times dS_T  
&=&  \int_{-\infty}^{\infty} \Lambda(\nu)
\times \frac{A^{'}_{T}\exp(\sigma_T(\xi-\xi_0))}{\left(1+\frac{\xi^2}{\nu}\right)^{\frac{\nu+1}{2}}}\times d\xi \\\nonumber
&\ge& A_T\times \int_{-\infty}^{x_c} \Lambda(\nu)
\times \frac{\exp(\sigma_T\xi)}{\left(1+\frac{\xi^2}{\nu}\right)^{\frac{\nu+1}{2}}}\times d\xi \\\nonumber
&+& A_T \times \exp(\sigma_T x_c) \times  \int_{x_c}^{\infty} \Lambda(\nu)
\times \frac{d\xi}{\left(1+\frac{\xi^2}{\nu}\right)^{\frac{\nu+1}{2}}} .
\end{eqnarray}
	Let $Z$ be the value of the integral in the middle row of Eq. (6) and let $(1-p)$ be the value of the integral of the last line of Eq. (6), i.e., $(1-p) = P\{\xi > x_c\}$.  Equation (6) can then be rewritten as
\begin{equation}
\label{eq:7}
S_0\times \exp(r\times T) = A_T\times Z + A_T  \times \exp(\sigma_T x_c)\times (1-p)\ ,
\end{equation}
and solved for the value of $A_{T }$:
\begin{equation}
\label{eq:8}
A_T= \frac{S_0\times \exp(r\times T)}{Z+(1-p)\times\exp(\sigma_t x_c)}\ .
\end{equation} 
	For a truncated distribution where the pdf is set $= 0$ for $\xi > x_{c }, A_T = S_0\times \exp(r\times T) /Z$ and the pdf in the integral that defines $Z$ (middle line of Eq. (6)) will have a factor of $1/p$ where $p = P\{\xi \le  x_c\}$ is the probability that $\xi \le x_c$.  This
follows from the definition of conditional probability:  $P\{x < \xi \le x+dx \vert \xi \le x_c \} = P\{x < \xi \le x+dx~ \cap  \xi \le x_c \}/P\{\xi \le x_c \} = pdf(\xi)/p$ if $ \xi \le x_c$ and zero if $\xi > x_c$. 

	Note that $Z = Z(\sigma, x_{c }(p), t, \nu)$ and that $Z$ is not a function of the random variable $\xi$. 

	For a log normal distribution with $\mu = 0$ and $\sigma = 1, x_c = +\infty, p = 1, Z =\exp(\sigma_T^2/2)$,
 and $A_T = S_0\times \exp(r\times T - \sigma_T^{2}/2)$.  Thus $C_T = S_0\times \exp(r\times T)\times N( (\ln(K_T /S_0)
- r\times T - \sigma_T^{ 2}/2)/\sigma_T ) - K_T\times N( (\ln(K_T /S_0) - r\times T + \sigma_T^{ 2}/2)/\sigma_T )$ where $N(x)$ is the
cumulative distribution function (CDF) for the unit normal pdf, $N(x) = P\{X \le  x\vert \sigma=1\}$. 
 
	$C_0 = C_T\times \exp(-r\times T) = S_0\times N(d_1) - \exp(-r\times T )\times K \times N(d_2)$ where $d_1 = (\ln(S_0/K_T) + r\times T + \sigma_T^{ 2}/2)/\sigma_T$ and $d_2 = d_1 - \sigma_T$.   This is the standard form for the Black-Scholes formula for a European call  \cite{bs73,Hull2006}.  Note that $K_0 = K_T \exp(-r\times T)$ and the expressions can be rewritten in terms of $K_0: C_0 = S_0\times N(d_1) - K_0\times N(d_2)$ with $d_1 =
(\ln(S_0/K_0) + \sigma_T^{ 2}/2)/\sigma_T$ and $d_2 = d_1 - \sigma_{T }$. 
 
	Thus the value for $A_T$ appears correct and it should be possible to price an option assuming a Student's {\it t}-distribution (or any other distribution) using the approach and equations developed in this manuscript.  
 
	The price at the time of expiration for a European call, $C_T$, using a Student's {\it t}-distribution and capping the value of the asset at a critical value $x_c$ is 
\begin{eqnarray}
\label{eq:9}
C_T &=& \Lambda(\nu) \times  \int_{\frac{\ln(K_T/A_T)}{\sigma_T}}^{x_c} \frac{A_T\exp(\sigma_T\xi)-K_T}{\left(1+\frac{\xi^2}{\nu}\right)^{\frac{\nu+1}{2}}}\times d\xi \\\nonumber
&+& \Lambda(\nu) \times \left(A_T \exp(\sigma_T x_c)-K_T\right)\times  \int_{x_c}^{\infty}  \frac{d\xi}{\left(1 +\frac{\xi^2}{\nu} \right)^{\frac{\nu+1}{2}}}\ .
\end{eqnarray}
	For a capped asset, $A_T$ is defined by Eq. (8). 

	For a truncated {\it t}-distribution, the price of a call $C_T$ is obtained by dropping the last line of Eq. (9), dividing the integral in the first line of Eq. (9) by $p(x_c) = P\{\xi \le x_c\}$, and using $A_T = S_0\times \exp(r\times T) /Z$.  It is helpful to remember that $Z$ is defined by some of the integral on the first line of Eq. (9) (see Eqs (7) and (6) for the definition of Z) and thus is also divided by a factor of $p(x_c)$ for a truncated pdf.  The value of the integral in the last line of Eq. (9) equals $p(x_c)$, which we shorten to $p$. 

	The price for a European put option at time $T$ is $P_T$ and equals
\begin{equation}
\label{eq:10}
P_T = \Lambda(\nu) \times
 \int_{-\infty}^{\frac{\ln(K_{T }/A_{T })}{\sigma_T}} \frac{K_T - A_{T }\exp(\sigma_{T }\xi)}{\left(1+\frac{\xi^2}{\nu}\right)^{\frac{\nu+1}{2}}}\times d\xi \ ,
\end{equation}
for both a capped asset and a truncated pdf, where the truncation occurs at $x_c > \ln(K_T/A_T)/\sigma_T$.  The value of $Z$ is different for the capped and truncated approaches, and this makes $A_T$ different for the two approaches.  

	In pricing a put, one could also truncate the distribution (i.e., set the lower limit of integration to a critical value $x_p$ and set the pdf to zero for values $< x_p$) or use a floor at the critical value (rather than a ceiling as was done for a call).  The values of $Z$ and $A_T$ would be modified in a straightforward manner to accomodate the floor or truncation.  In this paper we have chosen to use the full range of the integral in Eq. (10).  There are six possibilities to calculate an option: two choices for the call (cap or truncate \textemdash one of which must be chosen) and three choices for the put (floor, truncate, or nothing).  Our goal is to present the approach to pricing an option using a Student's $t$-distribution.  In this regard, is it necessary to truncate or to cap the distribution at the upper end.  We leave discussion of the merits of a floor and truncation to future work.  

	We consider a formula for the price of an European option that is obtained using either a maximum value for the asset or a truncated Student's {\it t}-distribution to be a Gosset formula. 

	There is little difference in philosophy between options priced with a capped value or with a truncated pdf.  In both approaches the return is assumed to be always less than a critical value.  With the capped asset approach, more weight is applied to the maximum value of the asset than for the truncated pdf approach.  The truncated pdf is more compact to write as there is not a second, additive term.  There is, however, a difference in risk between the two approaches.  The truncated pdf approach assumes that the value of the asset will never exceed the critical value.  There is no guarantee that the value will not exceed the critical value. 

	The approach of capping the payout of a benefit is common in insurance.  Capping the payout reduces the cost to purchase the benefit.  Capping the payout removes the risk associated with the benefit exceeding the critical value.  It is known, by design, that the payout will never exceed the cap but it is accepted that there is a non-zero probability that the expense will exceed the benefit. 

	Capping the value of the asset underlying a call option removes the risk associated with the value of the asset exceeding a critical value (assuming that the buyer of the call option agrees to the condition that the payout not exceed the cap value), and thus reduces the cost of the option.  Capping the value is one approach to keep the integrals, which are needed to price an option, finite.  The capped Gosset formula would be the proper formula to use to price a European option for an asset that is capped in value. 

	Put-call parity holds regardless of the underlying pdf.  Consider $E\{S_T - K_T\} = E\{ ( S_T-K_T)^{+} - (K_T- S_T)^{+}\} = C_T - P_T$ where $(x)^{+} = max\{x, 0\}, C_T$ is the cash value of a call evaluated at time $T$, and $P_T$ is the cash value of a put evaluated at time $T$. 

	For a martingale (a fair wager), $E\{S_t\} = S_0\times \exp(r\times t)$, and thus $E\{S_T - K_T\} = S_0\times \exp(r\times T) - K_T = C_T - P_{T }$.  From the time value of money, $x_t = x_0\times \exp(r\times t)$, then $C_0 - P_0 = S_0 - K_T\times \exp(-r\times T)$ or $C_0 - P_0 = S_0 - K_0$. 

	One of the confusing features of pricing an option are the time values.  Typically the option is priced at time $= 0$ for a stock price at time $= 0$ and a strike price at time $= T$.  One is usually given $S, \sigma$, and $r$ implicitly for time $= 0$ (i.e., $S_0$, and $r = r(0)$), the strike price $K$ implicitly for time $T$, (i.e., $K = K_T$ ), and asked implicitly for $C$ and $P$ at time $=0$ (i.e., $C_0$ and $P_0$).  The volatility $\sigma$ is evaluated at time $= 0$, and is propagated to time $= T$ using diffusion: $\sigma_T = \sigma\times \sqrt{T}$.  For
diffusion, the variance increases linearly with time.  $\sigma_T$ is used in the pricing of the options.  For pricing an option with the Gosset formula, it is necessary to specify the shape parameter $\nu$, which is a statement of belief about the range of values that the volatility might take at time = $T$. 

\section{Comparison of Prices from Gosset and Black-Scholes Formulae}

	Figure 4 is comprised of plots of the price of a call option calculated using the Gosset formula minus the price of the same option calculated using the Black-Scholes formula, as a function of the number of degrees of freedom assumed for the Student's {\it t}-distribution.  A capped distribution was used for the Gosset formula and confidence levels $p$ of 0.99, 0.999, and 0.9999 were used.  For small numbers of the degrees of freedom $\nu$ and $p = 0.9999$, the Gosset formula predicts a price difference $>$ \$5 between the values predicted by the Gosset and Black-Scholes formulae.  For the calculations, a risk-free rate $r = 3\%$, $S_0 = \$50.00$, $K_T = \$49.00$, $\sigma_T = 0.3$, and $T = 1.0$ 
year were assumed.  The Black-Scholes formula predicts a price of \$7.12 for a call option with these parameters.  The Student's {\it t}-distribution is considered to be equivalent to the normal distribution for the number of degrees of freedom $\nu > 40$.  The differences between the prices obtained from the Gosset formula and the Black-Scholes formula are between \$0.06 and \$0.11 for 40 degrees of freedom for confidence levels of 0.99, 0.999, and 0.9999. 

	Figure 5 is comprised of similar plots as for Fig. 4, except that a truncated distribution was used to calculate the price of the call option with the Gosset formula.  The price of a call option obtained by using a truncated distribution is less than the price obtained by using a capped distribution.  For a confidence level of 99\% and $\nu > 25$, the Gosset formula predicts a value for a call option that is less
than the value predicted by the Black-Scholes formula.  This is not surprising given that for large $\nu$, the Student's \textit{t}-distribution approaches the normal pdf.  As the two distributions become similar, any approach that excludes data in the tails will predict a lower value for the option. 
\begin{figure}[hb]
\begin{minipage}[t]{0.45\linewidth}
\centering
\includegraphics[scale=0.4]{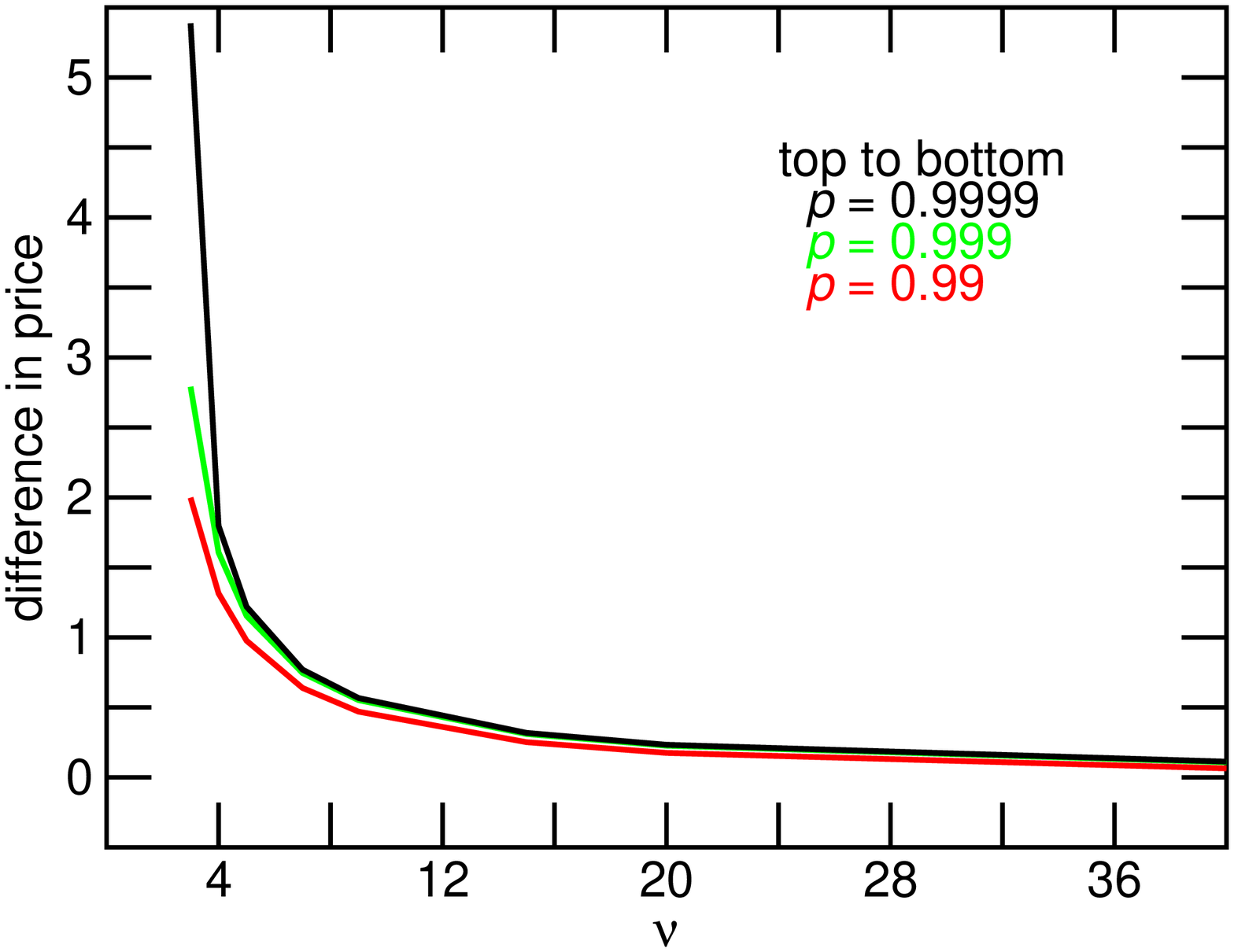}
\caption{Price difference, Gosset $-$ Black-Sholes, as a function of $\nu$ for confidence levels of 0.99, 0.999, and 0.9999 using a capped distribution.}
\label{fig:figure4}
\end{minipage}
\hspace{0.6cm}
\begin{minipage}[t]{0.45\linewidth}
\centering
\includegraphics[scale=.4]{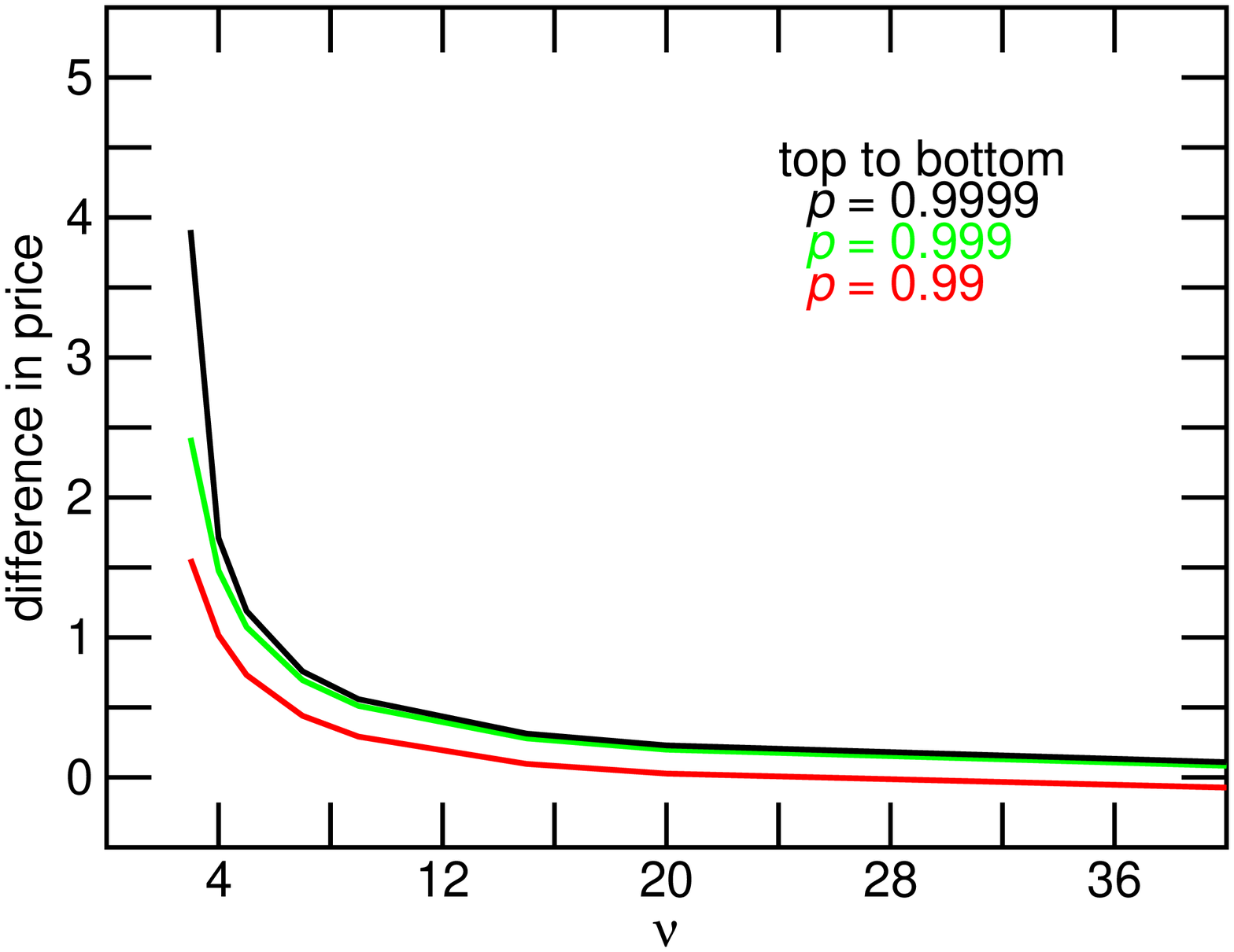}
\caption{Price difference, Gosset $-$ Black-Sholes, as a function of $\nu$ for confidence levels of 0.99, 0.999, and 0.9999 using a truncated distribution.}
\label{fig:figure5}
\end{minipage}
\end{figure}
The difference in cost of the option between the capped asset approach and the truncated distribution approach owes to contributions from three terms.  The $Z$ term is different for the two approaches and this leads to differences in the integrands and lower limits of integration.  For $p = 0.9999$, $\nu = 3$, $r = 0.03$, $\sigma_T = 0.3$, $T = 1.0$, $S_0 = \$50.00$, and $K_T = \$49.00$, the price difference, capped minus truncated, is \$1.48.  The contribution to cost from the values above the cap (i.e., the contribution from the last line of Eq. (9)) contributes \$3.14 to the difference.  The integrand for the truncated distribution is larger than for the capped asset.  This contributes $-$ \$1.52 to the difference.  The lower limits of integration are slightly different for the capped asset ($Z = 1.281$ yielding $\ln(K_T/A_T)/\sigma_T = 0.6583$) and for the truncated distribution ($Z = 1.203$ yielding $\ln(K_T/A_T)/\sigma_T = 0.4488$).  This area contributes $-$\$0.10, for a total difference of \$1.52 in the costs of the call option when calculated using the two approaches. 

	Figure 6 gives the value of a call option using the capped Gosset formula minus the value of a call option computed using the Black-Scholes formula, as a function of the confidence level.  For large numbers of degrees of freedom $\nu$, the Gosset formula predicts a value that is less than the value predicted by the Black-Scholes formula.  Again, this is not surprising as the Student's {\it t}-distribution will approach the normal pdf for large $\nu$ and any pricing mechanism that excludes area in the tails will predict a lower price.

	Figure 6 demonstrates that the price of an option for a \textit{t}-distribution with a small $\nu$ increases dramatically as the confidence level increases.  For a confidence level approaching 100\%, the price tends to infinity since one of the integrals required to price the option diverges.  Although we use the term confidence level $p$, it is helpful to remember that the confidence level gives a critical value, $x_c(p)$.  The exponential of the critical value, $\exp(\sigma_{T }x_c )$, gives the value that the asset will not exceed in a capped asset approach, or is hoped will not be exceeded in a truncated pdf approach. 

	Figure 7 presents similar information as Fig. 6 and is presented for completeness.  The data for Fig. 7 was calculated using a truncated pdf rather than a capped value for the asset. 

\begin{figure}[h]
\begin{minipage}[t]{0.45\linewidth}
\centering
\includegraphics[scale=0.4]{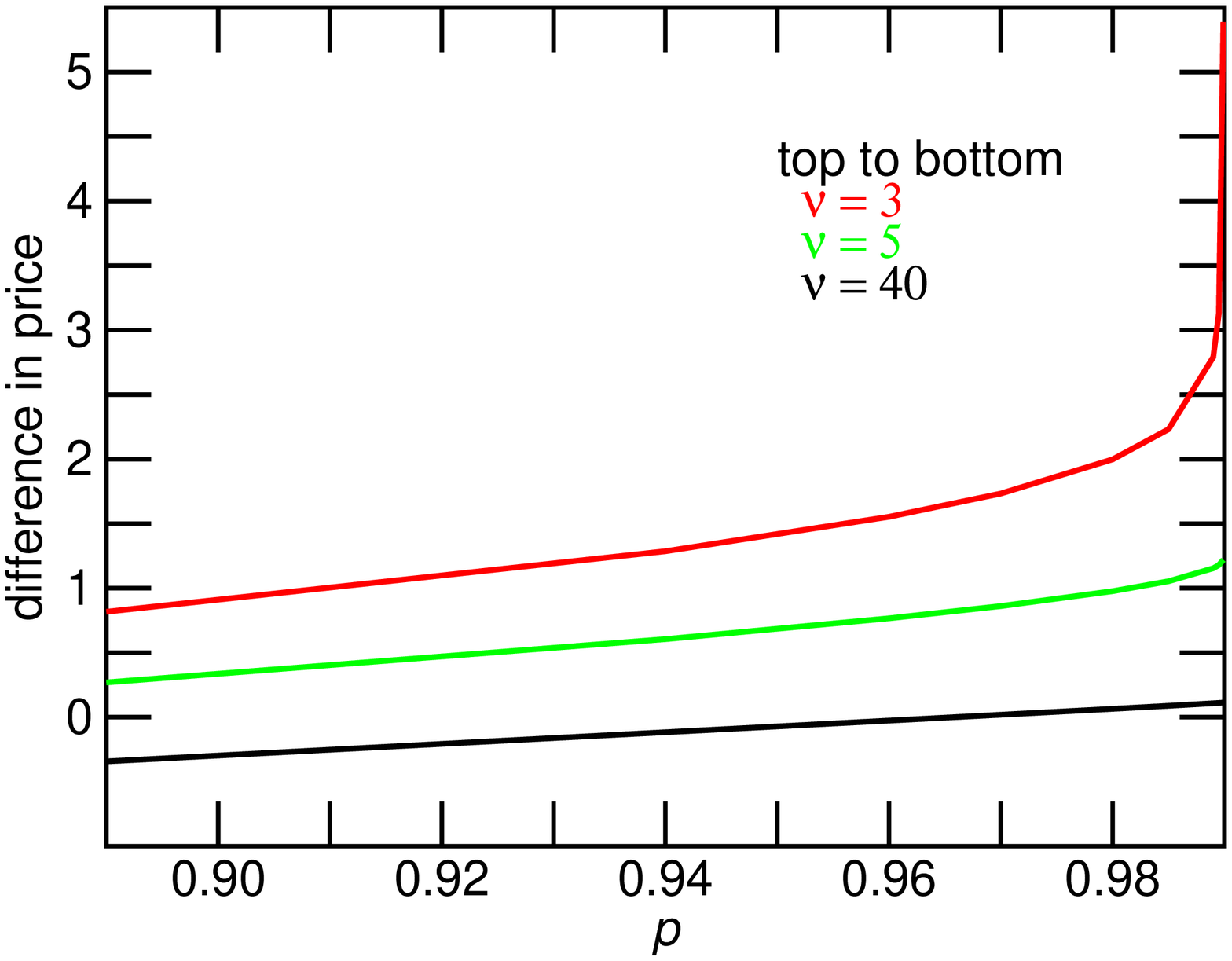}
\caption{Price difference, Gosset $-$ Black-Sholes, of a call option as a function of the confidence level for $\nu$ = 3, 5, and 40 for a capped distribution.}
\label{fig:fig6}
\end{minipage}
\hspace{0.6cm}
\begin{minipage}[t]{0.45\linewidth}
\centering
\includegraphics[scale=.4]{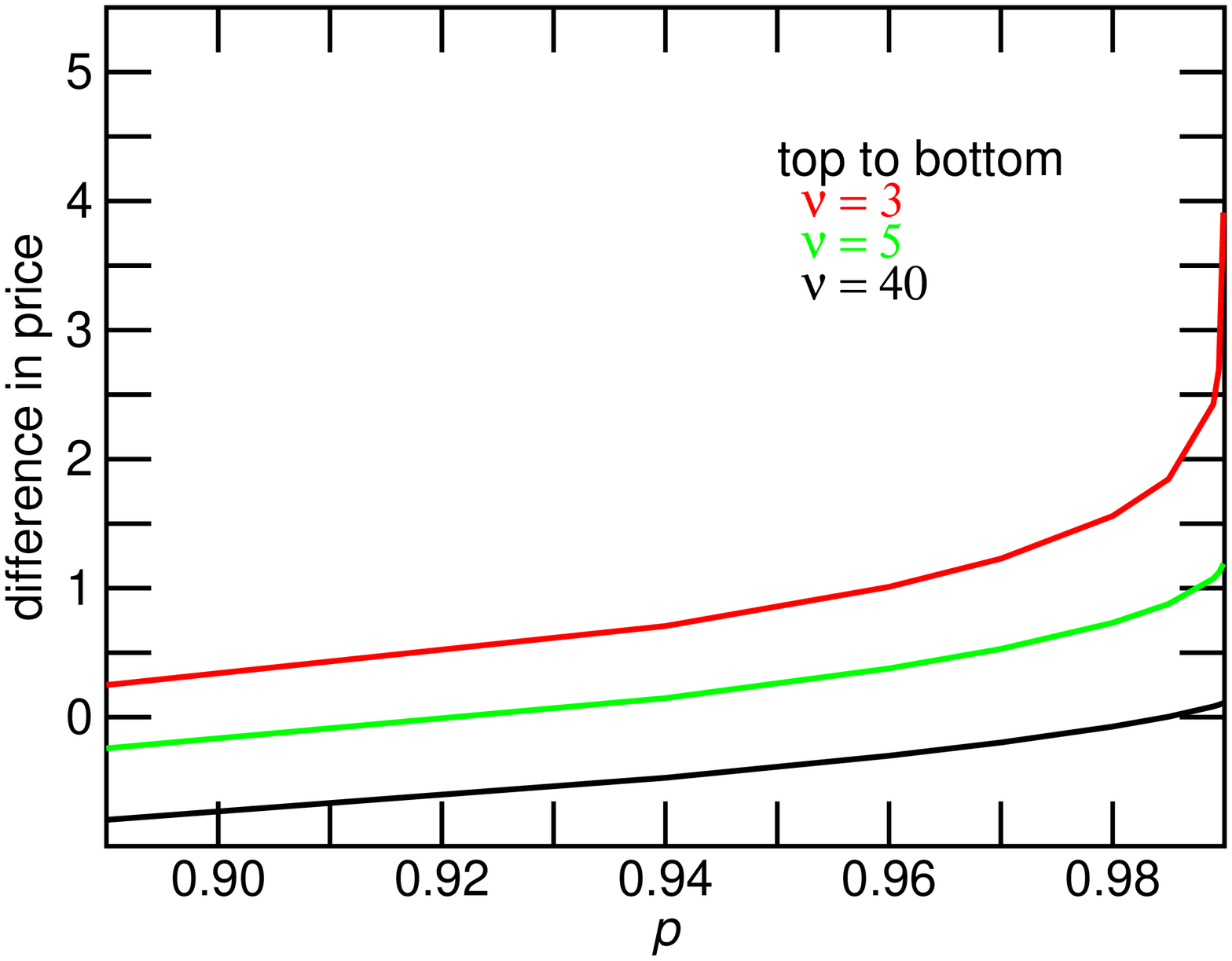}
\caption{Price difference, Gosset $-$ Black-Sholes, of a call option as a function of the confidence level for $\nu$ = 3, 5, and 40 for a truncated distribution.}
\label{fig:fig7}
\end{minipage}
\end{figure}

	Figure 8 gives the prices of calls and puts as a function of the $t = 0$ price of the asset. The capped Gosset formulae for $C_T$ and $P_T$, Eqs. (9) and (10), were used to calculate the prices of the options.  When the option prices were used in the put-call parity formula, the difference was of order of the precision of the calculation.  This held for both the truncated and capped approaches.  Figure 8 displays prices for $\nu = 3$, $\nu = 5$, and for the Black-Scholes formula.  A confidence level of 99.9\%, $K_T$ = \$49.00, $r = 3\%$, $\sigma_T = 0.3$, and $T = 1.0$ year were used.  The curve for $\nu  = 40$ essentially overwrites the Black-Scholes curve for the scale used.  The difference between the $\nu = 40$ and Black-Scholes prices were less then \$0.10 for the range of $S_0$.   

	Figure 9 gives the prices of calls and puts as a function of the $t = 0$ price of the asset. The truncated Gosset formulae for $C_T$ and $P_T$, Eqs. (9) and (10), were used to calculate the prices of the options.  A comparison of Figs. 8 and 9 shows that the truncated Gosset formula predicts lower prices for the options than does the capped Gosset formula.  The truncated Gosset formula ignores events in the tails and thus costs less than the capped Gosset formula, which allows for events in the tails but places a limit on the value of the events in the tails.

\begin{figure}[h]
\begin{minipage}[t]{0.45\linewidth}
\centering
\includegraphics[scale=0.4]{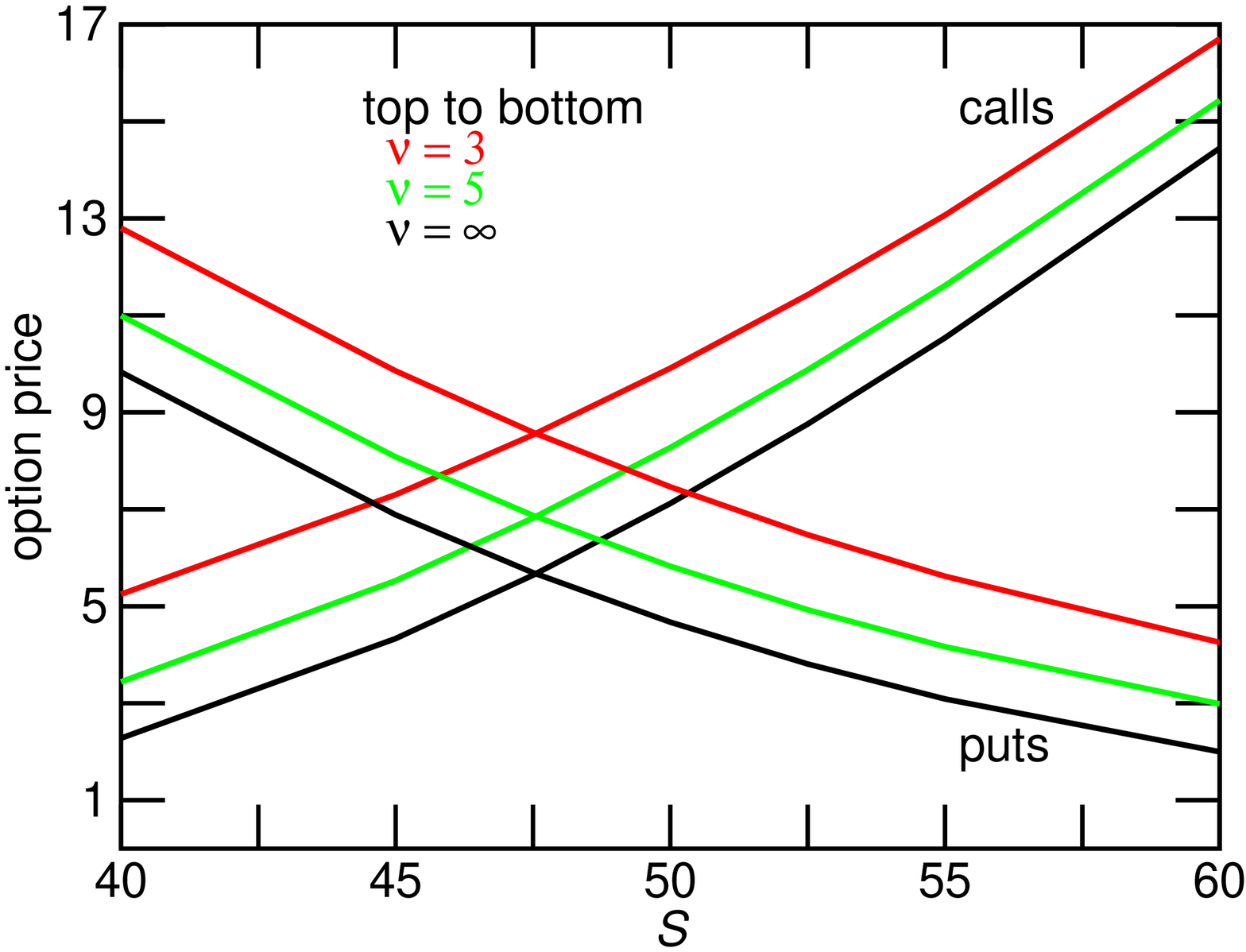}
\caption{Prices of put and call options for $\nu = 3$ and $\nu= 5$ using a capped Gosset formula with $p=0.999$, and for the Black-Scholes formula.}
\label{fig:fig8}
\end{minipage}
\hspace{0.6cm}
\begin{minipage}[t]{0.45\linewidth}
\centering
\includegraphics[scale=.4]{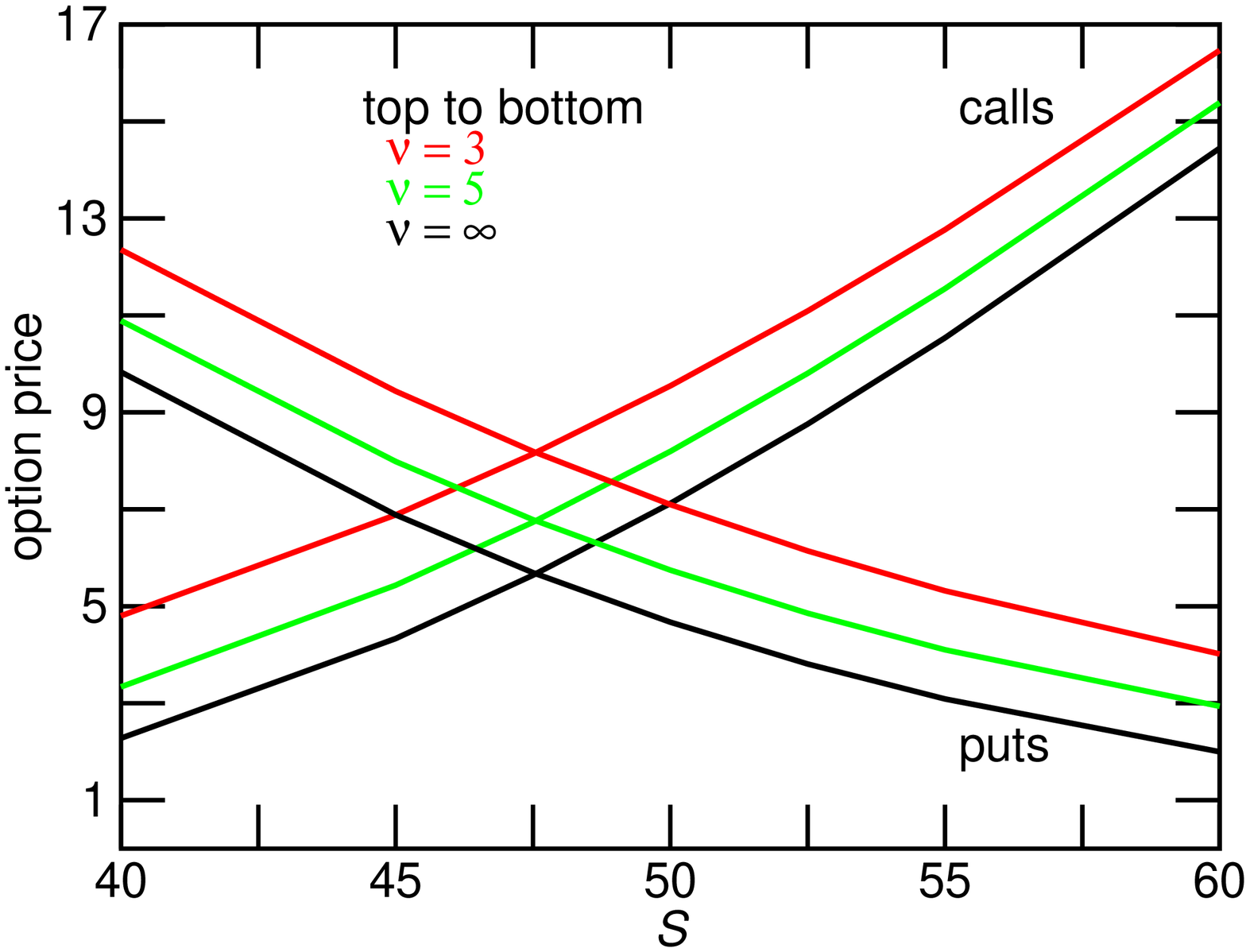}
\caption{Prices of put and call options for $\nu = 3$ and $\nu= 5$ using a truncated Gosset formula with $p=0.999$, and for the Black-Scholes formula.}
\label{fig:fig9}
\end{minipage}
\end{figure}

\section{Conclusion}

	In this paper we have presented a method to price European options using a Student's {\it t}-distribution.  We refer to this method as a Gosset method and the formulae for pricing options as Gosset formulae, in honour of the person who published under the {\it nom de plume} (or pseudonym) Student.  The Gosset formula essentially extends the Black-Scholes formula by removing the constraint that the volatility be constant in time. 

	Student's {\it t}-distributions have \textquotedblleft fat tails\textquotedblright\space and fit the returns from assets better than a normal pdf.  The difficulty with pricing options using the Student's {\it t}-distribution is that the fat tails cause one of the integrals, which is needed to price an option, to diverge.  A viable and physically reasonable value for the price of the option can be obtained by either capping the value of the underlying asset or by truncating the distribution.  The value for the cap or where the truncation takes place need not be unrealistic.  The required integral can be evaluated for a confidence level of 99.99\%, which implies a price increase of $>$ 4 000\% for realistic parameters (volatility $\sigma_T = 0.4$ and shape parameter $\nu = 5$) of the Student's {\it t}-distribution.  An increase of 4 000\% in the value of an asset would be phenomenal and perhaps unphysical. 

	There is little difference in philosophy between the approaches of pricing an option by capping the value of the asset or truncating the pdf.  Both approaches reduce the cost of the option.  However, there can be a large difference in the exposure to risk between the two approaches. 

	Truncation of the pdf assumes that the value of the underlying asset will never exceed the value where the truncation occurs.  There is no guarantee that this value will not be exceeded. 

	Capping the value of the asset ensures, by design and with consent, that the risk associated with the value of the option exceeding the value of the cap does not exist.  In the cap approach, it is accepted that the value of the asset might exceed the value of the cap.  However, the price of the option is set assuming that the payout will not exceed the value of the cap.  Provided that the option writer and buyer agree to this approach, then there is no risk associated with a large increase in value of the asset.  The buyer gives away the right to extreme returns in exchange for a lower price for the option. 

	If the option writer and buyer do not agree to a cap on the payout, then the approach of capping the value of the asset still holds risk that the value of the asset will exceed the cap value.  For the same critical value and parameters for the pdf, the cap approach offers an increased premium for the risk as compared to the approach of truncation of the pdf.  This is because the cap approach allows for the possibility that the value of the asset will exceed the value of the cap. 

	The Student's {\it t}-distribution has an additional parameter (the number of degrees of freedom, $\nu$) as compared to the normal pdf.  This parameter is a statement of the belief about the range of values that the volatility might take at the time that the option expires.  A large value of $\nu$ is a statement that the volatility will change little in time.  In the limit as $\nu$ approaches infinity, the Student's $t$-distribution approaches the normal distribution and hence the Gosset formula for the price of an option becomes the same as the Black-Scholes formula.  The Black-Scholes formula is derived on the assumption that the volatility is constant in time and known.  A small value of $\nu$ is a statement that the volatility might take one of a broad range of values.  Fits to the DJIA for returns over 100 years give a value of $\nu = 2.65$.  This value of $\nu =2.65$ allows for the distribution of volatility that has been observed over the 100 years of returns.   In turbulent markets, it would be wise to use a small value for the number of degrees of freedom and a large confidence level.  A confidence level is required in the pricing of options using a Student's $t$-distribution.  In calm markets, it might be acceptable to use a large $\nu$, of order $N-1$ where $N$ is the number of independent samples that are used to calculate the volatility, and a smaller confidence level to match the expectations of the future and tolerance to risk.  The flexibility offered by the Gosset formulae of choice of approach (cap or truncate), of the number of degrees of freedom to match the fat tails of returns, and of the critical value allow the analyst and investor additional opportunity to use their insight and circumstances to price options. 

	The approach used to derive the Gosset formulae can be adapted to distributions other than the Student's {\it t}-distribution.  This may be of interest in situations and for returns on assets that do not appear to be well described by a Student's {\it t}-distribution.

\begin{acknowledgments}
	This work was funded in part by the Natural Sciences and Engineering Research Council of Canada. 
\end{acknowledgments}

\end{document}